\documentclass[preprint,12pt, nofootinbib]{revtex4-1}

\usepackage{amssymb,amsmath,setspace,graphicx,mathrsfs, float,slashed, mathrsfs, amsfonts, setspace, natbib, gensymb, siunitx, gensymb, hhline, booktabs, multirow}

\usepackage[usenames, dvipsnames]{color}

\usepackage[bottom]{footmisc}

\begin{document}
	
\title{Leptoquark solution for both the flavor and ANITA anomalies}

\date{\today}

\author{Bhavesh Chauhan}
\email{bhavesh@prl.res.in}
\affiliation{Physical Research Laboratory, Ahmedabad 380009, India}
\affiliation{Indian Institute of Technology Gandhinagar, Palaj 382355, India}

\author{Subhendra Mohanty}
\email{mohanty@prl.res.in}
\affiliation{Physical Research Laboratory, Ahmedabad 380009, India.}

\begin{abstract}
The ANITA experiment has seen anomalous Earth emergent showers of EeV energies which cannot be explained with Standard Model interactions. In addition, tests of lepton flavor universality in $ R(D^{(\ast)})$ and $R(K^{(\ast)})$ have shown significant deviations from theoretical predictions. It is known that, among single leptoquark solutions, only the chiral vector leptoquark $U_1 \sim (\mathbf{3}, \mathbf{1} , 2/3)$ can simultaneously address the discrepancies. In this paper, we show that the leptoquark motivated by flavor anomalies coupled to a sterile neutrino can also explain the ANITA Anomalous Events. We consider two scenarios, (a) the sterile neutrino, produced via resonant leptoquark mediated neutrino-nucleon interactions, propagates through the Earth without significant attenuation and decays near the surface to a $\tau$ lepton; and (b) a cosmogenic sterile neutrino interacts with the matter near the surface of Earth and generates a $\tau$ lepton. These two scenarios give significantly large survival probabilities even when regeneration effects are not taken into account. In the second scenario, the distribution of emergent tau energy peaks in the same energy range as seen by ANITA. 
\end{abstract}

\maketitle



\section{Introduction}

The ANtarctic Impulsive Transient Antenna (ANITA) instrument is designed to detect interaction of ultra-high energy neutrinos via the Askaryan effect in ice. During its first and third flight, it also observed unexpected upward directed showers apparently emerging well below the horizon \cite{Gorham:2016zah, Gorham:2018ydl, Schoorlemmer:2015afa}. The observed signal is consistent with $\tau$ induced showers. The essential details of the two Anomalous ANITA Events (AAEs) are given in Table \ref{tab:1}.  The survival probability ($\epsilon$) is estimated taking into account the neutrino regeneration effects and $\tau$ energy losses in \cite{Ref-Fox} using only Standard Model (SM) interactions.
\begin{table}[h]
	\centering
	\begin{tabular}{|l | c | c|}
		\hline
		\textbf{Property} & \textbf{AAE1}  & \textbf{AAE2} \\
		\hline
		Energy ($E_\tau$) & $0.6 \pm 0.4$ EeV & $0.56^{+0.3}_{-0.2}$ EeV\\
		Zenith Angle  & $117.4 \pm 0.3$ \degree & $125.0 \pm 0.3$ \degree\\
		Chord Length ($l_{\oplus}$) & $5740 \pm 60$ km & $7210 \pm 55 $ km \\
		$\epsilon_{SM}$ & $4.4 \times 10^{-7}$ & $3.2 \times 10^{-8}$ \\
		\hline
	\end{tabular}
	\caption{\label{tab:1} Properties of the anomalous events.}
\end{table}

The small survival probabilities within SM indicate that new physics scenarios should be invoked to explain these events. In the past, the AAEs have been explained in the framework of sterile neutrinos \cite{Ref-Huang, Cherry:2018rxj}, Supersymmetry \cite{Ref-Fox, Ref-Bhu, Anchordoqui:2018ssd}, and CPT symmetric universe \cite{Anchordoqui:2018ucj}. However, each of these explanation have their own limitations \cite{Ref-Bhu}. Similarly, collider experiments such as LHCb, Belle, and BaBar have observed hints of Lepton Flavor Universality Violation (LFUV) in semi-leptonic decays of the B meson. In particular, the ratios
\begin{equation}
R(D^{(\ast)}) = \frac{\mathcal{BR}(\bar{B}\to D^{(*)}\tau^-\bar{\nu}_{\tau})}{\mathcal{BR}(\bar{B}\to D^{(*)}\ell^-\bar{\nu}_{\ell})},
\end{equation}
\begin{equation}
R(K^{(\ast)}) = \frac{\mathcal{BR}(\bar{B}\to \bar{K}^{(*)}\mu^+\mu^-)}{\mathcal{BR}(\bar{B}\to \bar{K}^{(*)}e^+e^-)}.
\end{equation}
where $ \ell = e, \mu$ are known to have very weak dependence on hadronic form factors and provide excellent probes of LFUV \cite{Hiller:2003js}. The experimentally measured value of the observables $ R(D^{(\ast)})$ \cite{Huschle:2015rga, Hirose:2016wfn} and $  R(K^{(\ast)})$ \cite{Aaij:2017deq, Aaij:2017vbb} is consistently below SM prediction and together are dubbed as 'flavor anomalies' in this paper. These discrepancies can be explained in several extensions of SM, for example with leptoquarks \cite{Bhattacharya:2014wla, Bauer:2015knc, Ref-DG, Crivellin:2018yvo, Buttazzo:2017ixm,Biswas:2018snp, Assad:2017iib, Fajfer:2015ycq, Sakaki:2013bfa,Calibbi:2017qbu, Crivellin:2017zlb, ColuccioLeskow:2016dox, Blanke:2018sro, Chauhan:2017uil, Ref-Bes}.  \\

It was proposed in \cite{Ref-Fox} that a long lived BSM particle, which is produced in ultra-high energy (UHE) neutrino nucleon interactions, propagates freely through the chord of Earth, and decays to a $\tau$ near the surface can explain the AAEs. A natural candidate for this is $\tilde{\tau}$ (stau) in R-Parity conserving Supersymmetry \cite{Ref-Fox} and neutralino (mostly Bino) in R-Parity violating Supersymmetry \cite{Ref-Bhu}. In this paper, we consider two scenarios wherein a leptoquark, proposed as a resolution to flavor anomalies, also explains AAEs. In the first scenario, we extend the minimal leptoquark model of \cite{Buttazzo:2017ixm} with a heavy sterile neutrino ($\chi$). The SM singlet $\chi$ is produced in UHE neutrino-nucleon interactions mediated by the leptoquark. The sterile neutrino can travel inside Earth without significant attenuation and decays near the south pole. One of the decay products is the $\tau$ particle whose shower is seen by ANITA. In the second scenario, an cosmogenic UHE sterile neutrino propagates freely through the chord of the Earth and produces a $\tau$ via leptoquark mediated interaction. Interestingly, the same leptoquark interaction also explains $R(D^{(\ast)})$ through $b \rightarrow c \tau \chi$ as shown in \cite{Ref-DG}.\\

In Sec. II, we estimate the number of AAEs for isotropic and anisotropic flux. In Sec. III, we provide details of the leptoquark model and discuss the two scenarios in detail before we conclude in Sec. IV. 

\section{ANITA Anomalous events}


In order to estimate the number of Earth emergent showers seen by ANITA, we evaluate the survival probability $\epsilon$ (also called efficiency in \cite{Ref-Huang}) which represents the fraction of incident flux $\Phi$ that is converted into $\tau$ near the surface. We use the expression
\begin{equation}
\label{noe}
\mathcal{N} =  A \cdot \delta T \cdot \delta \Omega   \int_{E_{min}}^{E_{max}}  d E_\nu  \cdot \epsilon \cdot \Phi(E_\nu)
\end{equation}
where the effective area of ANITA $A \approx 4~ km^2$ is estimated using the Cherenkov angle \cite{Ref-Huang}, $\delta T$ is the time period, and $\delta \Omega$ is the acceptance angle. For temporally continuous source, $\delta T \approx 25 $ days is the combined exposure of ANITA-I (17.25 days) and ANITA-III (7 days) \cite{Gorham:2016zah, Gorham:2018ydl}. We have ignored the contribution of ANITA-II (28.5 days) as it was not sensitive to such events. For transient sources, $\delta T$ will depend on the source and can be smaller. For isotropic source, $\delta \Omega \approx 2 \pi~sr$. However, for anisotropic source, 
\begin{equation}
\delta \Omega \approx 2 \pi (1 - \cos \delta_\theta)~\approx~0.0021~sr
\end{equation}
where $\delta_\theta \sim 1.5 \degree $ is the angular uncertainty relative to parent neutrino direction \cite{Gorham:2018ydl}. The neutrino energy ($E_\nu$) is integrated over the range which gives correct range of shower energy. For example, if $\tau$ is produced through interaction of the incident neutrino such that $E_\tau = E_\nu / 4$. Since the observed shower has energy in the range 0.1 - 1 EeV, one must integrate over 0.4 - 4 EeV.  In general, $\epsilon$ depends on $E_{\nu}$ and model parameters.

We now provide order-of-magnitude estimate of the required $\epsilon$ taking $\delta T = 25$ days. For the isotropic case, we assume that the source of EeV neutrinos is the Greisen-Zatsepin-Kuzmin (GZK) mechanism. We approximate the GZK flux by the upper limit of its saturated value over the range 0.4 - 4 EeV as, 
\begin{equation}
\overline{\Phi}_{iso}\approx 10^{-25}~(\text{GeV cm}^2 \text{ s sr})^{-1}
\end{equation}
which gives $\mathcal{N} \approx 200 \epsilon$. To get two events, one requires $\epsilon \sim 0.01 $. Similar estimates were also obtained in \cite{Ref-Bhu} which takes energy dependence into account albeit with larger exposure time. With the Standard Model interactions, the authors in \cite{Ref-Fox} have estimated that $\epsilon_{SM} \sim 10^{-7}$ for the two reported events. Thus the estimated number of anomalous events from GZK neutrinos with only SM interactions is,
\begin{equation}
\mathcal{N}_{iso}^{SM} \sim  2 \times 10^{-5}
\end{equation}
which makes observation of two events extremely unlikely. \\

One can relax the assumption that the source of EeV neutrinos is the GZK flux. This allows us to postulate that such high energy neutrinos are coming from a localised source in the sky. The upper limit on such anisotropic flux of EeV neutrinos is, 
\begin{equation}
\overline{\Phi}_{aniso}\approx 3.2 \times 10^{-20}~(\text{GeV cm}^2 \text{ s sr})^{-1}
\end{equation}
which is several orders larger than the isotropic case. After accounting for the small solid angle one can similarly obtain, $\mathcal{N} \approx 2.1 \times 10^{4}~\epsilon$. To get two events, one requires $\epsilon \sim 10^{-4} $. Using SM interactions for the incident neutrinos,
\begin{equation}
\mathcal{N}_{aniso}^{SM} \sim  2.1 \times 10^{-3}
\end{equation}
which again makes the two events very unlikely. In this section, we have ignored the energy dependence of $\epsilon$ as well as $\Phi$. Even after taking those into account, the message will remain unchanged. The smallness of $\epsilon_{SM}$ makes the two event unlikely. \\

One must also check the compatibility of IceCube with ANITA observations. Even though IceCube has smaller effective area, the long duration of the experiment implies that the expected number of EeV scale up going $\tau$-tracks seen by IceCube ($\mathcal{N}_{IC}$) to be larger than expected anomalous events by ANITA ($\mathcal{N}_{AN}$). Using the relative exposures, it has been estimated that $ \mathcal{N}_{IC} \approx 10 \times \mathcal{N}_{AN}$ \cite{Ref-Huang, Ref-Bhu}. In \cite{Ref-Fox}, the authors identify three events in nine year (3142 days) IceCube data that may have origin similar to ANITA. This implies that $\mathcal{N}_{AN} = 0.3$. Using Poisson distribution, the probability of observing two such events is around 0.03.  The challenge for BSM scenarios is to get $\mathcal{N}_{AN}$ of this order by enhancing $\epsilon$ as has been done in the two scenarios studied in this paper.

\section{Leptoquark resolution of AAE}
As has been discussed \cite{Ref-Bes, Ref-DG}, a vector leptoquark $U_1$ with $SU(3)_C \times SU(2)_L \times U(1)_Y$ quantum numbers $(\mathbf{3}, \mathbf{1} , 2/3)$ can simultaneously explain the flavor anomalies. It is also one of the handful models that admit leptoquark coupling to a sterile neutrino \cite{Dorsner:2016wpm}. The interaction of $U_1$ with fermions in the mass basis is, 
\begin{equation}
\label{lag}
- \mathcal{L} \supset (V \cdot g_L )_{ij} ~\bar{u}_L^i \gamma^{\mu}U_{1,\mu} \nu_L^j +  (g_L)_{ij}~ \bar{d}_L^i \gamma^{\mu}U_{1,\mu} e_L^j +  (g_R)_{ij}~\bar{d}_R^i \gamma^{\mu}U_{1,\mu} e_R^j + (g_\chi)_i~ \bar{u}_R^i \gamma^\mu U_{1, \mu} \chi_R 
\end{equation}
where $V$ is the CKM matrix and contribution of PMNS matrix is ignored.

\subsection{Heavy Sterile Neutrino}

In this section, we assume that the sterile neutrino is sufficiently heavy so that its contribution to semi-leptonic B decays is kinematically forbidden. Even though the interaction of up-type quarks with a sterile neutrino can generate dangerous scalar and pseudo-scalar operators, their contributions can be neglected and the conclusions in \cite{Ref-Bes} remain unchanged. The required texture of coupling matrices is,
\begin{equation}
g_L = \begin{pmatrix}
0 & 0 & 0 \\
0 & g_{s\mu} & g_{s\tau} \\
0 & g_{b \mu} & g_{b \tau}
\end{pmatrix}  \quad 
g_R = 0. \quad 
g_{\chi} = (0~ g_{x} ~0).
\end{equation} 
The left-handed coupling ($g_L$) generates the desired Wilson coefficients (i.e. $\delta \mathcal{C}_9 = - \delta \mathcal{C}_{10}$ with the correct sign for $b \rightarrow s \mu \mu $ and $g_{V_L} > 0$ for $b \rightarrow c \tau \nu$). In this way, $U_1$ is one of the rare solutions that can simultaneously address both the anomalies. The right-handed coupling ($g_R$) is severely constrained as it generates scalar and pseudoscalar operators that are disfavored. The sterile neutrino $\chi$ can also couple to other up-type right handed quarks, but we have neglected those couplings and their constraints for simplicity. In this section, we will assume the mass of leptoquark $U_1$ to be $M_U = 1.5$ TeV and the couplings to be, 
\begin{equation}
\label{bmcoup}
g_{s\mu} = - 0.012, \quad g_{b\mu} = 0.2, \quad g_{s\tau} = 0.5 , \quad g_{b \tau} = 0.5
\end{equation}
which can explain the flavor anomalies. Such a choice is within the reach of future LHC searches but allowed from present limits \cite{Ref-Bes, Biswas:2018snp}. We treat the coupling $g_x$ and mass of the sterile neutrino ($M_\chi$) as free parameters of the theory. \\ 

\begin{figure}[h!]
	\centering
a	\includegraphics[width=6cm]{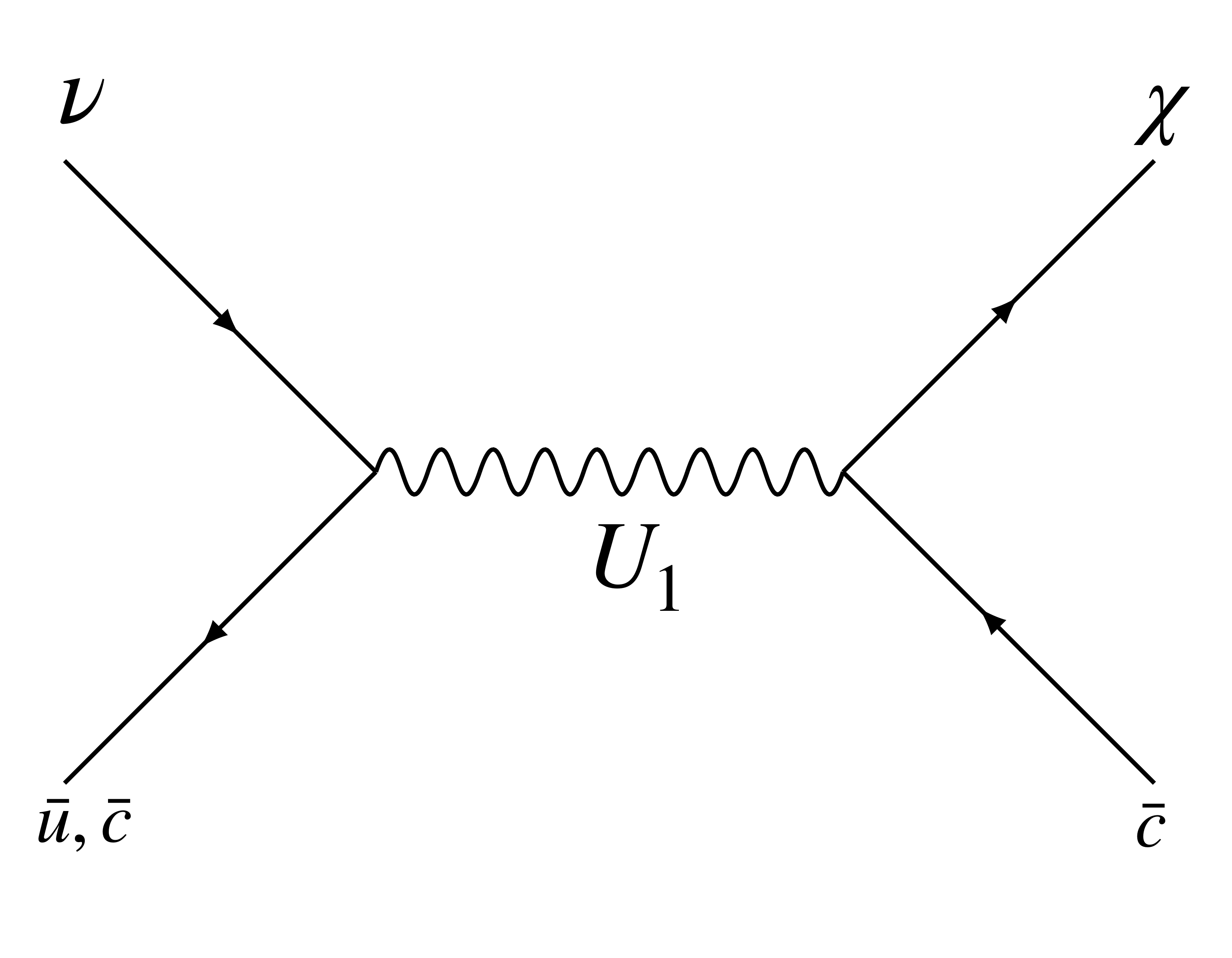}
	\includegraphics[width=6cm]{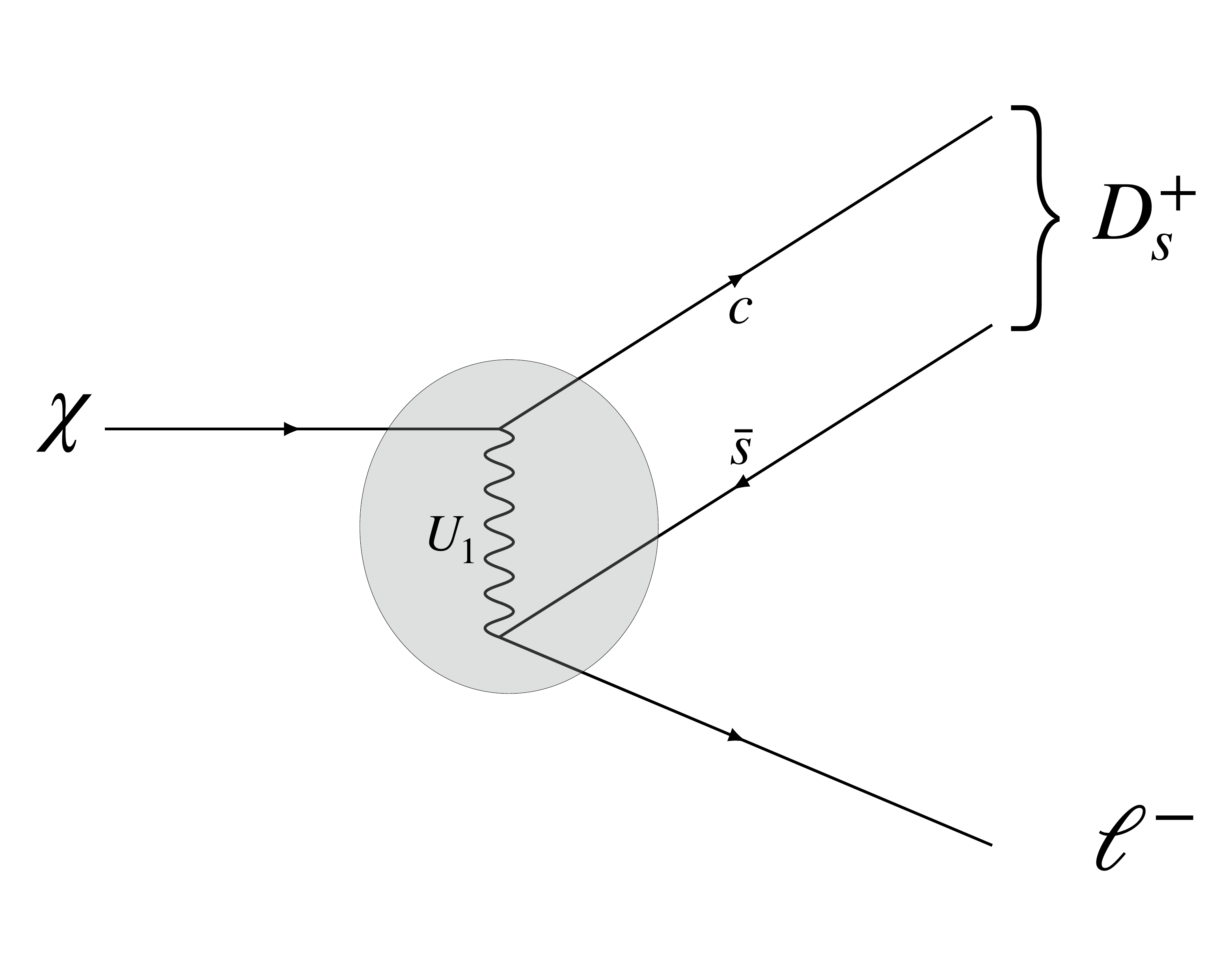}
	\caption{\label{fig:prod_decay} The Feynman diagrams for the process involved in Model A. \emph{Left:} The s-channel neutrino quark interaction mediated by leptoquark $U_1$ that produces sterile neutrino in final state is shown. \emph{Right}: The decay mode of sterile neutrino to charged lepton and $D_s^+$ is shown. The shaded circle represents the effective vertex.}
\end{figure}

The singlet is produced near the surface of Earth through neutrino-nucleon interaction mediated by the leptoquark. It is assumed that the cross section for the process is dominated by the resonant s-channel neutrino-quark interactions. It has been pointed out in \cite{Becirevic:2018uab} that the gluon initiated process can also give significant contributions. However, this will give an $\mathcal{O}(1)$ correction to survival probability and has been neglected for the heavy sterile neutrino case. The production cross section can be approximated in the narrow width limit as,
\begin{equation}
\sigma_{LQ} (E_\nu) =  \frac{3 \pi}{2} \left(  \frac{g_x^2}{g_x^2 + 1.08} \right) \frac{1}{ 2 M_N E_\nu} \int_{0}^{1} dy  y^2 \left( (0.11)^2 f_u  + (0.5)^2 f_c   \right)
\end{equation}
where $f_q$ is the parton distribution function (PDF) of $q$ evaluated at $x = M_U^2/ 2 M_N E_\nu$ and $Q = M_U \sqrt{y}$. We have used ManeParse \cite{Clark:2016jgm} and NNPDF3.1(sx) \cite{Ball:2017nwa, Ball:2017otu} datasets for the PDFs. The numerical factors (1.08, 0.11, and 0.5) are obtained using the central value of CKM parameters \cite{Tanabashi:2018oca}. Note that the PDFs are evaluated at small-$x$ where the quark and anti-quark PDFs are similar and hence neutrino and anti-neutrino have similar cross section. The interaction length is estimated as, $\ell_{LQ}=(\rho N_A \sigma_{LQ})^{-1}$ where we have used $\rho \approx 4 $ and $N_A = 6.022 \times 10^{23} \text{ cm}^{-3}$ in water-equivalent units. Even though the density is larger near the center of Earth, the approximation for density is valid for the chord lengths relevant for AAE. \\ 

As opposed to previous studies with three body decay of a singlet \cite{Ref-Bhu}, in this paper we estimate the two body decay width of the sterile neutrino to a pseudoscalar meson and the tau lepton. Since the decay width is being estimated in the rest frame of sterile neutrino of mass few GeV, one can integrate out the heavy leptoquark and write the effective Lagrangian as, 
\begin{equation}
\label{Leff}
\mathcal{L}_{eff} = \frac{2 g_{x} g_{q\ell}}{M_{U}^2} \left[ \bar{c} P_L q \right] \left[ \bar{\ell} P_R \chi \right]
\end{equation}
where $q \in \{s,b\}$ and $\ell \in \{\mu, \tau\}$. We also use the expression, 
\begin{equation}
\label{formfac}
\langle 0 | \bar{q}_1 \gamma_5 q_2 | P \rangle = i \frac{M_P^2}{M_1 + M_2} f_P
\end{equation}
where $P$ is a pseudoscalar meson of mass $M_P$ and $f_P$ is the associated form factor. The rest frame partial width of the sterile neutrino is, 
\begin{equation}
\Gamma_\tau \equiv \Gamma(\chi \rightarrow \tau^- D_{s}^+) =  \frac{1}{16 \pi} \left( \frac{g_x g_{s\tau}}{M_U^2} \right)^2 \left(\frac{M_{D_s^+}^2}{M_c + M_s} f_{D_s^+}\right)^2 M_\chi ~\beta\left( M_{D_s^+}, M_\tau, M_\chi \right)
\end{equation}
where the phase space factor is, 
\begin{equation}
\beta(a,b,c) = \left[ \left( 1 - \left(\frac{a - b}{c}\right)^2\right)\left( 1 - \left(\frac{a + b}{c}\right)^2\right)\right]^{1/2}.
\end{equation}
For numerical estimation we use, 
\begin{equation}
	f_{D_s^+} = 257.86 ~\text{MeV}  \quad 	M_{D_s^+} = 1.968 ~\text{GeV} \quad
\end{equation}
and the quarks and lepton masses used are $M_c = 1.29$ GeV, $M_s = 95$ MeV, $M_\mu = 105.66$ MeV, and $M_\tau = 1.77$ GeV respectively. The associated decay length of $\chi$ in Earth's frame is estimated as, 
\begin{equation}
\ell_{D} = \gamma c \tau = \frac{1}{\Gamma_\tau}\frac{E_\chi}{M_\chi} \approx \frac{1}{\Gamma_\tau}\frac{E_\nu}{2 M_\chi}.
\end{equation}
where the last approximation is true for the range of energies involved. In this scenario, $E_\tau = E_\nu/4$ and hence for shower energy $\sim 0.5$ EeV, one requires the incident neutrino to have energy $E_\nu \sim 2$ EeV. \\

With only SM interactions, one can estimate the \emph{bare} survival probability $\epsilon_{0} = e^{- l_{\oplus}/\ell_{0}}$ where $l_{\oplus}$ is the length of path traversed by neutrino inside Earth and for EeV neutrinos, $\ell_{0} \sim 275 $ km \cite{Ref-Fox}. However, this is severely modified when one takes neutrino regeneration effects during propagation. In \cite{Ref-Fox}, the probability is obtained using simulations and mentioned in Table I. We denote these probabilities with $\epsilon_{SM}$. Due to the additional leptoquark interactions, the survival probability of the neutrino flux can be estimated using, 
\begin{equation}
\label{surv3}
\epsilon_{LQ} = \int_{0}^{l_{\oplus}} dl_1  \int_{l_{\oplus} - l_1 -  \delta}^{l_{\oplus} - l_1 }  dl_2  \left[   \frac{e ^{ - l_2 / \ell_{D}}}{\ell_{D}}  \frac{ e^{- l_1 / \ell_{LQ}}}{\ell_{LQ}}  \left(1 - \int_{0}^{l_1} \frac{dl_3}{\ell_{0}}e^{- l_3 / \ell_{0}} \right)  \right]
\end{equation}
The above expression can be understood as follows. The parentheses denote the fraction neutrinos that survives SM interactions after propagating a distance $l_1$. These neutrino undergo leptoquark interactions with the matter and produce a sterile neutrino. The sterile neutrino propagates a distance of $l_{\oplus} - l_1 - \delta$ before it decays near the surface of Earth in the $\delta \approx 10$ km window that will produce the observed $\tau$.  \\

\begin{figure}[h!]
	\includegraphics[width=7cm]{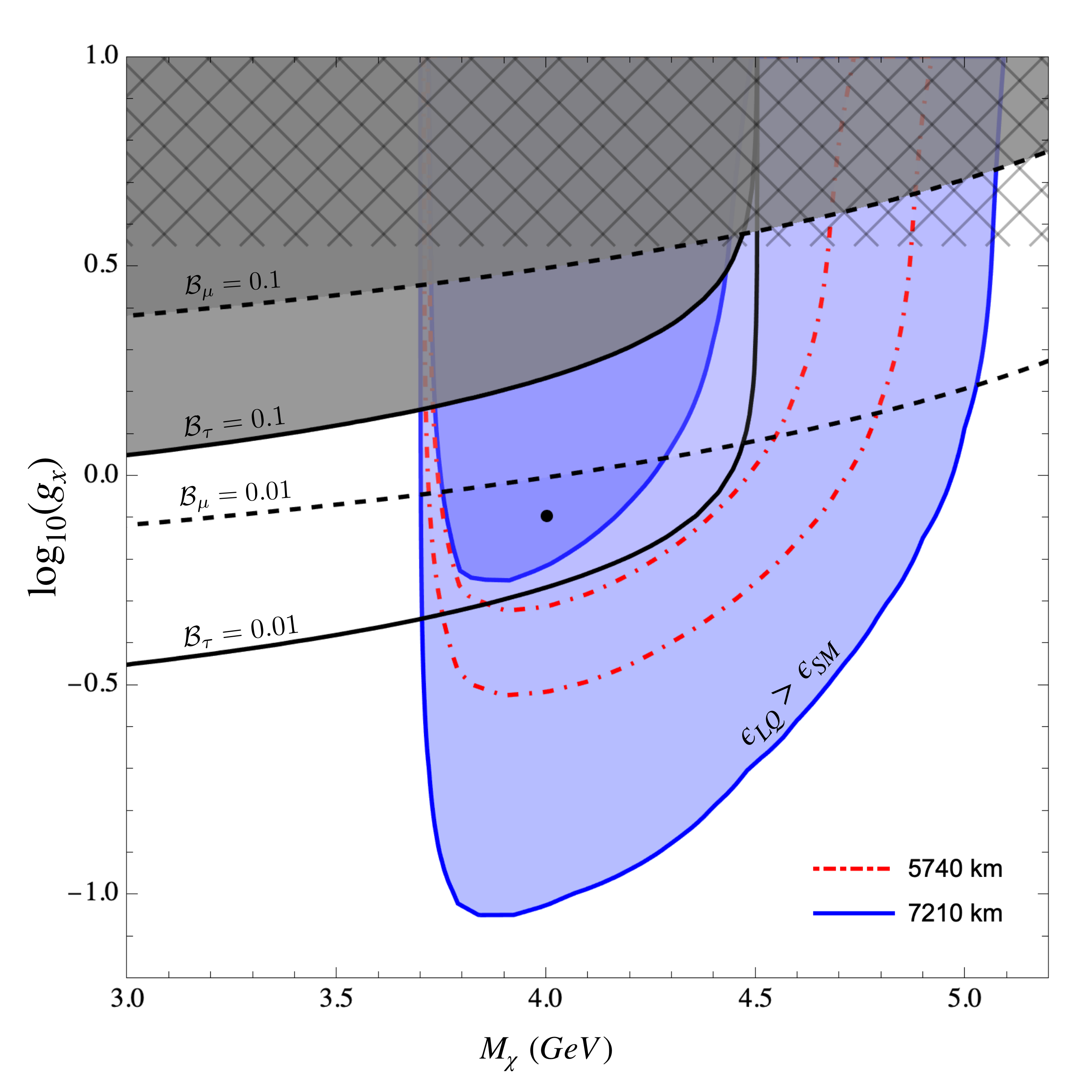}
	\caption{\label{fig:A} The parameter space that gives $\epsilon_{LQ} > \epsilon_{SM}$ (\emph{blue}), and $\epsilon_{LQ} > 1 \times 10^{-6}$ (\emph{dark blue}) for $l_{\oplus} =$ 7210 km is shown. Similar projections for $\ell_{\oplus} =$ 5740 km is shown by red curves. The gray shaded region is conservatively ruled out from $B_c^+$ decays and the limits for various $\mathcal{B_\ell}$ are shown. The top part is excluded using the perturbativity limit $g_x \leq \sqrt{4 \pi}$. The neutrino energy is fixed to be 2 EeV. The benchmark point considered in the text is shown.}
\end{figure}

In Fig. \ref{fig:A} we have shown the parameter space that gives $\epsilon_{LQ} > \epsilon_{SM}$, and $\epsilon_{LQ} > 1 \times 10^{-6}$ for the two values of $l_{\oplus}$. We find that the maximum survival probability in this scenario is of the order $4 \times 10^{-6}$. It is understood that neutrino regeneration effects can dramatically increase $\epsilon_{LQ}$ similar to SM. However, complete estimation requires simulation of neutrino propagation which is beyond the scope of this work. Moreover, we find that the precision measurement of $B_c^+$ decay modes can probe the most interesting part of the parameter space. We evaluate the branching fraction $\mathcal{B}_\ell = Br(B_c^+ \rightarrow \ell^+ \chi)$ for $\ell \in \{\mu,\tau \}$ to be, 
\begin{equation}
\mathcal{B}_\ell =  \frac{\tau_{B_c^+}}{4 \pi M_{B_c^+}} \left( \frac{g_x g_{b\ell}}{M_U^2} \right)^2 \left(\frac{M_{B_c^+}^2}{M_c + M_b} f_{B_c^+}\right)^2\left( M_{B_c^+}^2 - M_\ell^2 - M_\chi^2 \right)~\beta\left( M_\chi, M_\ell, M_{B_c^+} \right)
\end{equation}
where $f_{B_c^+} = 0.43$ GeV \cite{Ref-DG} and $M_{B_c^+} = 6.275$ GeV \cite{Tanabashi:2018oca}.  Since the typical branching ratio of leptonic mode is very small, we take the conservative limit of $ \mathcal{B}_\ell = 10\%$ for both $\mu$ and $\tau$ modes to constrain our parameter space. We also show limits for $ \mathcal{B}_\ell = 1\% $ which will be accessible in future B-factories and can test the model. \\

In this model, for the parameter space that we are interested in, the only kinematically allowed choice for the final state meson is $D_s^+$. The model also allows for $\chi \rightarrow \mu^- D_s^+$ however this decay mode is suppressed due to smallness of $|g_{s\mu}| \sim 0.012$ as compared to $|g_{s \tau}| \sim 0.5$ as seen in Eq. \eqref{bmcoup}. We also have $\chi \rightarrow \nu^- X$ but to get emergent $\tau$  one needs to account for another interaction in \eqref{surv3} which makes it less probable. This mode will be important when regeneration effects are evaluated using simulation and is beyond the scope of this paper. \\

To estimate the number of events, we consider the benchmark scenario 
\begin{equation}
M_{\chi} = 4.0~\text{GeV} \quad g_x = 0.8 
\end{equation}
for which $\Gamma_\tau = 4.64 \times 10^{-16}$ GeV and the survival fraction is $\epsilon_{LQ} \sim (1.5-2.0) \times 10^{-6}$. This gives the expected number of AAE per direction to be 0.03 using the saturated anisotropic flux. \\

In this scenario, larger values of the coupling $g_x$ seem to be preferable. However, they would be constrained from future measurements of $\mathcal{B}_\mu$. One can avoid these constraints if $g_{b\mu} = 0$ but then, the model cannot explain $R(K^{(\ast)})$. If one is willing to give up simultaneous explanation of both flavor anomalies, another interesting possibility opens up i.e. light sterile neutrino.

\subsection{Light Sterile Neutrino}

In \cite{Ref-DG}, it is shown that $U_1$ leptoquark coupled to a light sterile neutrino can also explain the flavor anomalies. However, as opposed to \cite{Ref-Bes}, $R(D^{(\ast)})$ is explained via right-handed couplings and $R(K^{(\ast)})$ via left-handed ones. It is seen that a simultaneous explanation in this scenario is in tension with big bang nucleosynthesis but $R(D^{(\ast)})$ can be explained successfully. The Lagrangian for the leptoquark is,
\begin{equation}
\label{lagB}
\mathcal{L}_{LQ} =  - \frac{1}{2} U_{\mu \nu}^{ \dagger}U^{\mu \nu} - i g_s \kappa  U_{\mu}^{ \dagger}T^a U_{\nu}G^{a \mu \nu} + M_U^2  U_{\mu}^{ \dagger} U^{\mu} + g_{b\tau} \bar{b}_R \gamma^{\mu}U_{1,\mu} \tau_R + g_x \bar{c}_R \gamma^\mu U_{1, \mu} \chi_R
\end{equation}
where $g_s$ is the strong coupling constant and $\kappa = 0 (1)$ for a minimally-coupled (gauge) theory. The excess can be explained with the following choice of coupling and leptoquark mass,
\begin{equation}
\label{eq:prodlim}
|g_x g_{b\tau} |\sim 0.62 \left( \frac{M_U}{1~\text{TeV}} \right)^2
\end{equation}
Considering the LHC constraints on the model, we chose $M_U = 1.5 $ TeV which is close to the lightest allowed mass for $\kappa = 1$. To a good approximation, $g_{b\tau} \in \{ 1.1, 1.4 \}$ which translates to $g_x \in (1.0, 1.25)$ using \eqref{eq:prodlim}. In this limit, the model has signatures in future 300 fb$^{-1}$ analysis. These limits are considerably weakened for $\kappa = 0$. One can refer to \cite{Ref-DG} for detailed discussion of the model and other constraints.\\  

\begin{figure} [h!]
	\includegraphics[width=14cm]{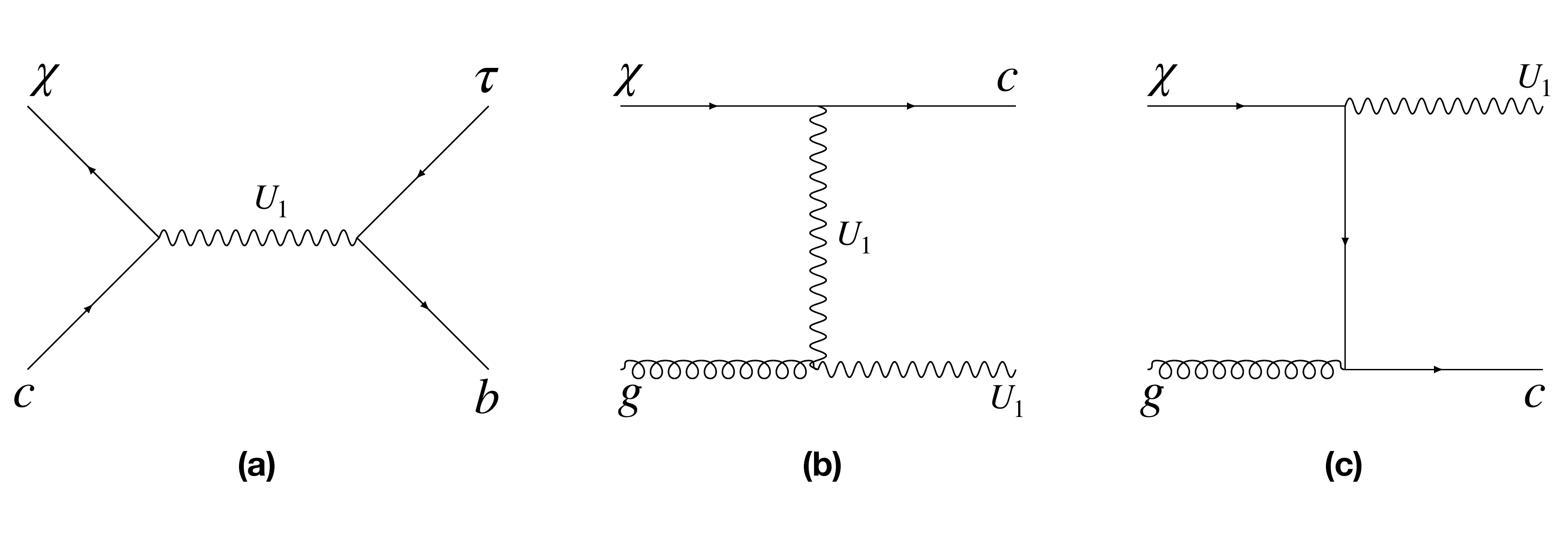}
	\caption{\label{fig:VV} The Feynman diagrams for $\chi$-nucleon interaction. (a) The dominant s-channel $\chi$-quark interaction. (b) The $\kappa$-dependent $\chi$-gluon interaction. (c) The $\kappa$-independent $\chi$-gluon interaction. }
\end{figure}

To explain AAE, we assume a flux of light sterile neutrinos ($\chi$) incident on Earth.  These sterile neutrinos can pass through the Earth almost unattenuated, however, a fraction of them can interact with the matter in Earth and produce a $\tau$ near the surface. In this section, we consider both $\chi$-quark and $\chi$-gluon interactions. The relevant Feynman diagrams are shown in Fig. \ref{fig:VV}  \\

\begin{figure} [h!]
	\includegraphics[width=9cm]{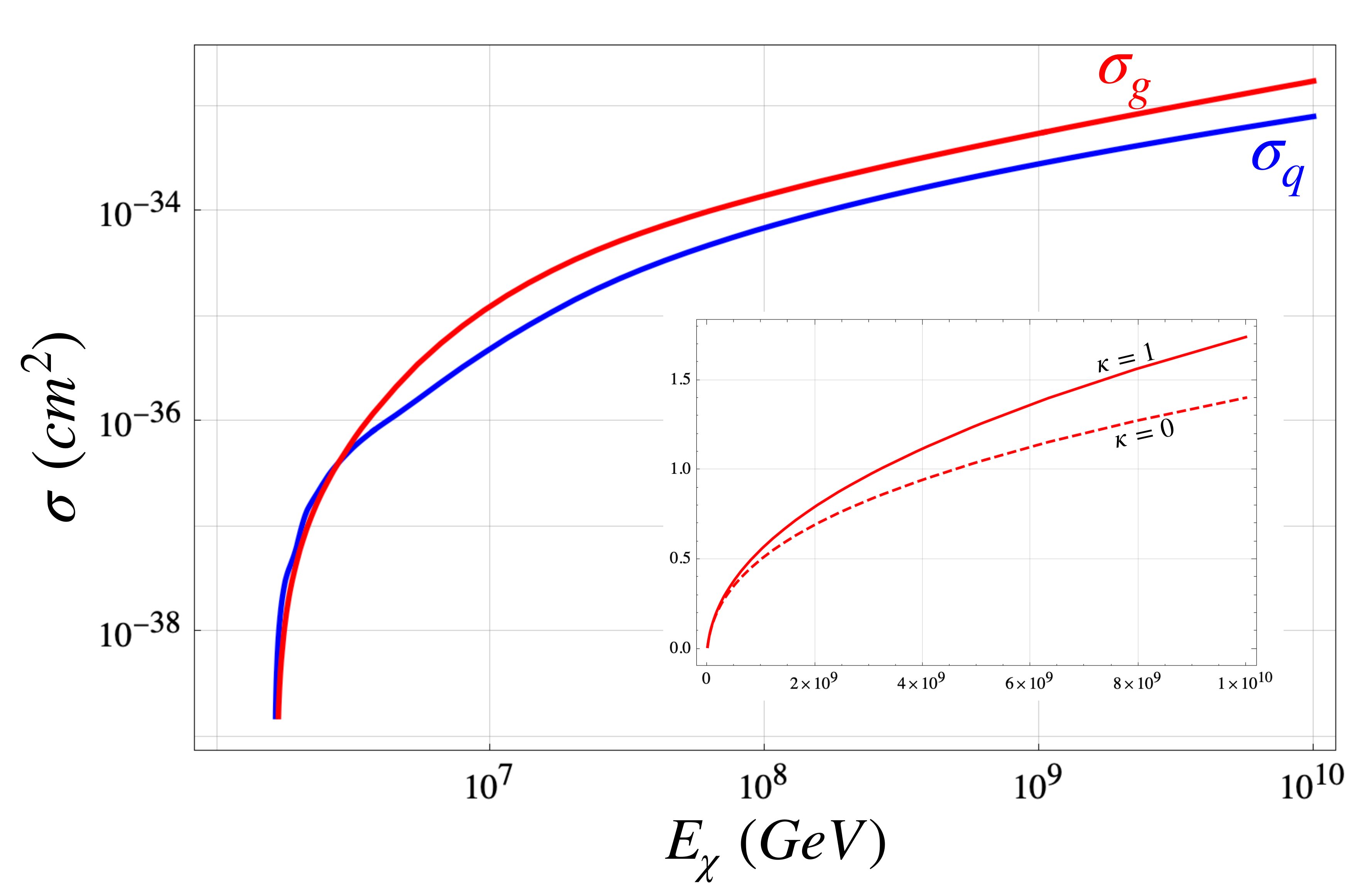}
	\caption{\label{fig:XS} The variation of cross section $\sigma_q~(\sigma_g)$ with incident sterile neutrino energy is shown in blue (red). The inset shows the difference in magnitude of $\sigma_g$ for $\kappa = $ 0 and 1 in arbitrary units.}
\end{figure}

The $\chi$-quark interaction is dominated by the s-channel resonant contribution and the cross section can be estimated by
\begin{equation}
\sigma_q = \sigma(\chi c \rightarrow \tau b) =  \frac{3 \pi}{2} \left(  \frac{g_x^2 g_{b\tau}^2}{g_x^2 + g_{b\tau}^2} \right) \frac{1}{ 2 M_N E_\nu} \int_{0}^{1} dy   (1 - y)^2  f_c .
\end{equation}
The difference in $y$- dependence is due to the $RR$ nature of interaction as opposed to $LR$ in the previous case. On the other hand, the $\chi$-gluon interaction cross section can be estimated using, 
\begin{equation}
\sigma_g = \sigma(\chi g \rightarrow \tau c \bar{b})  \approx  \sigma(\chi g \rightarrow c U_1) \times Br(U_1 \rightarrow \tau \bar{b}).
\end{equation}
We implemented the model in FeynRules \cite{Alloul:2013bka, Dorsner:2018ynv} and the cross section is calculated using CalcHep \cite{Belyaev:2012qa}. As was shown in \cite{Becirevic:2018uab}, the gluon initiated process are significant for large energies and of the same order of magnitude as the quark initiated processes. The cross section depends on $\kappa$ as evident from Fig. 3(b). In Fig. \ref{fig:XS}, we show the variation of $\sigma_q$ and $\sigma_g$ with incident sterile neutrino energy. We also show the relative strength for $\kappa =$ 0 and 1.

\begin{figure}[h!]
	\includegraphics[width=7cm]{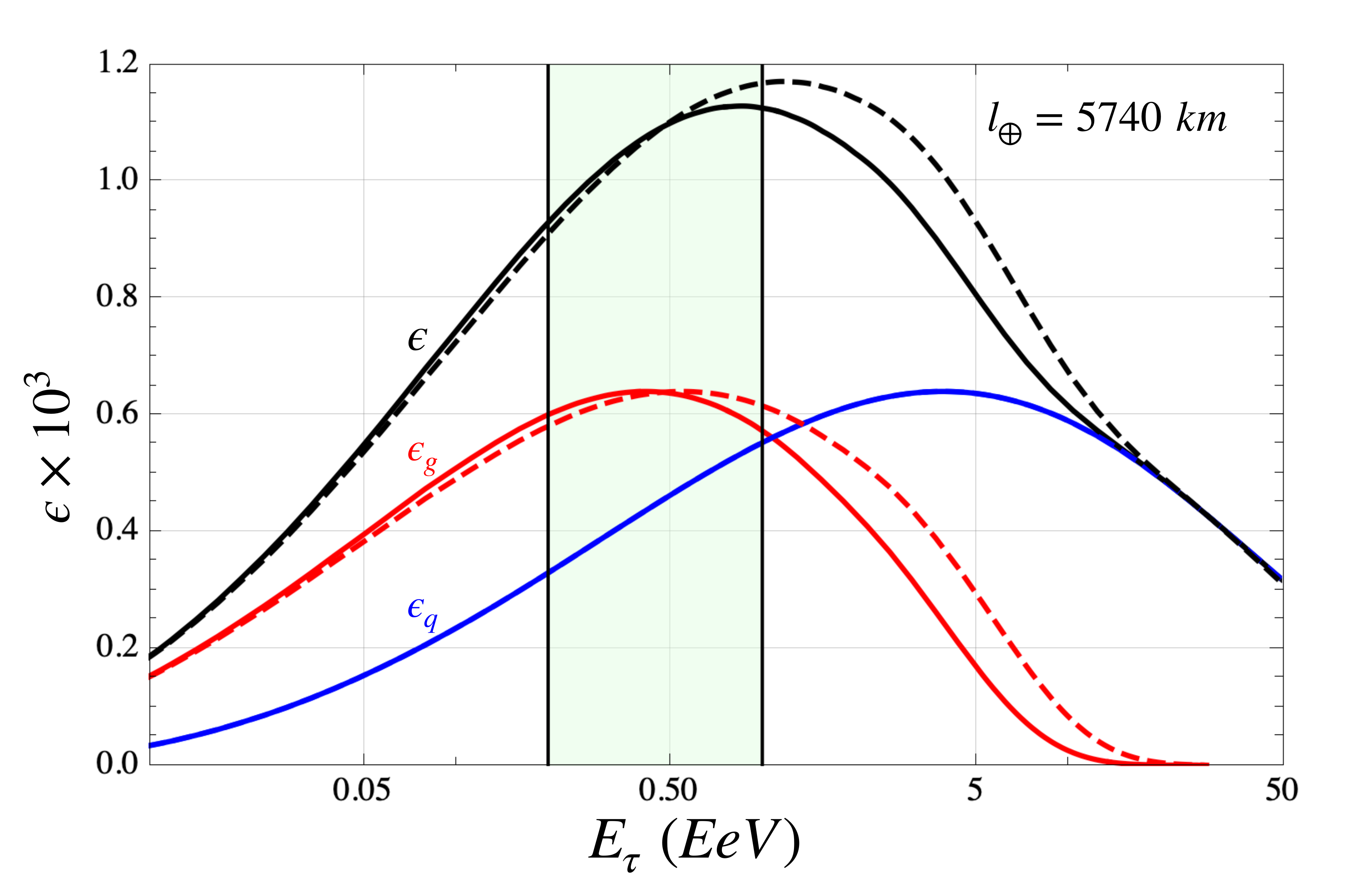} 	
	\includegraphics[width=7cm]{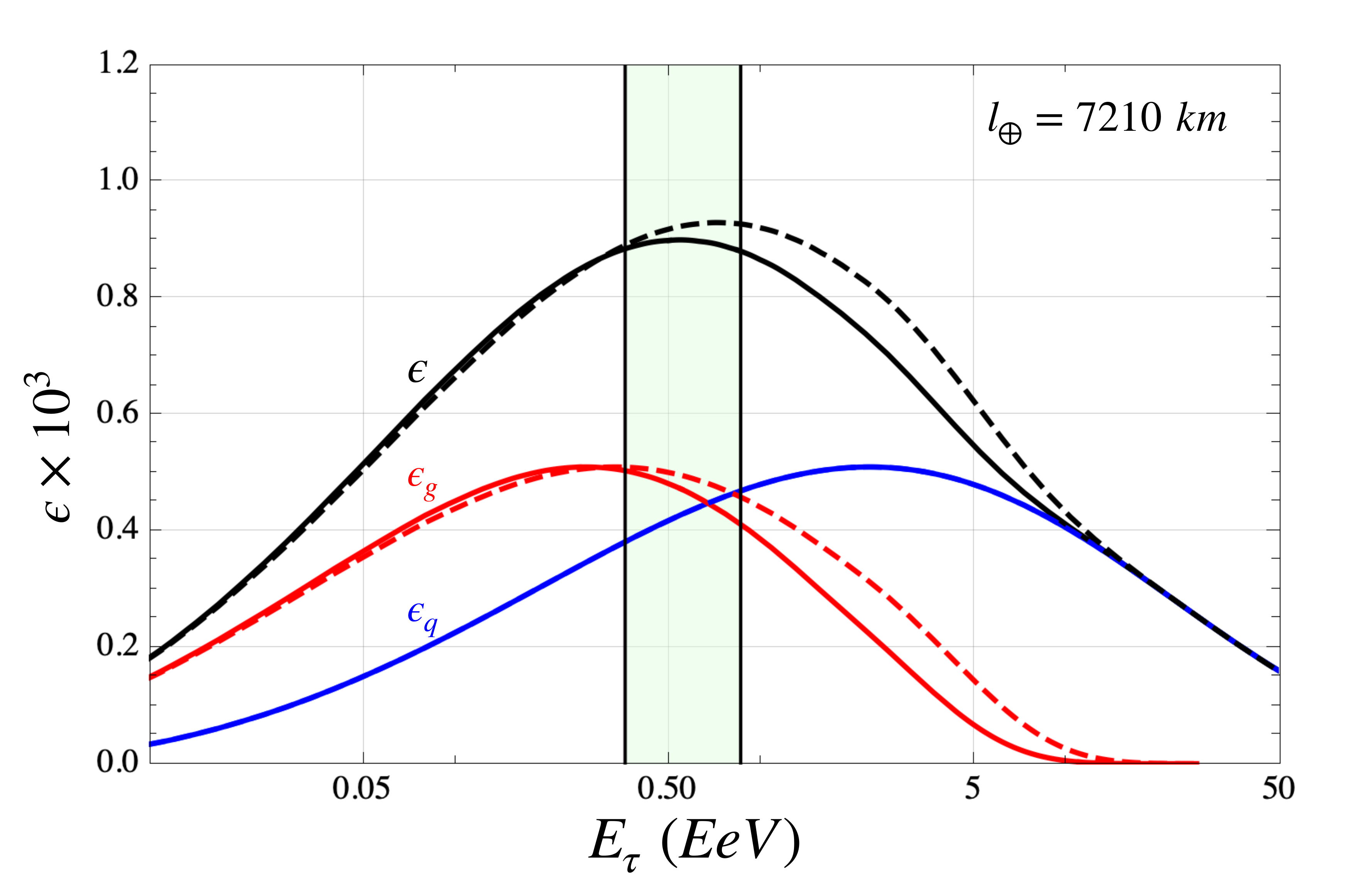}
	\caption{\label{fig:BRES} The variation of $\epsilon_{q}, \epsilon_g$, and $\epsilon$ is shown in blue, red, and black respectively. The solid curve is for $\kappa = 1$ and the dashed curve for $\kappa = 0$. The chord length $l_{\oplus}$ is fixed to be 5740 km (left) and 7210 km (right). We fix $g_x = 1.2$ for both the plots. The region shown in green is the observed shower energy for the two events.}
\end{figure}

The fraction of incident $\chi$ that interact with matter in Earth is given by, 
\begin{equation}
\label{survlight}
\epsilon_{q/g} = \int_{l_{\oplus} - \delta}^{l_{\oplus}} dl_1 \frac{ e^{- l_1 / \ell_{q/g}}}{\ell_{q/g}}
\end{equation}
where $\ell_{q/g} = (\rho N_A \sigma_{q/g})^{-1}$. One must note that, for $\chi$-quark interactions $E_\tau = E_\chi/2$ whereas for $\chi$-gluon interaction $E_\tau = E_\chi/4$. By uniformly varying $E_\chi$, we show the variation of $ \epsilon = \epsilon_{q} + \epsilon_{g}$ with energy of emergent tau in Fig. \ref{fig:BRES}. An interesting result of this scenario is that the distribution peaks for tau energy in the same range as seen by ANITA. \\

In order to estimate the number of events, one needs to know the flux of incident $\chi$ on Earth. It is clear from the discussion in Sec II. that this scenario cannot explain AAE with isotropic flux. We assume anisotropic flux from point-like sources in the sky and parametrize the incident flux as, 
\begin{equation}
\label{paramflux}
\Phi = \phi_0 \times 10^{-20} \left(\frac{E_\chi}{\text{EeV}}\right)^{- \gamma} ~ (\text{GeV cm}^2 \text{ s sr})^{-1}
\end{equation}
where the spectral index $\gamma$ is unknown. The number of events is then given by, 
\begin{equation}
\mathcal{N} \approx  \left(\frac{1800}{\text{EeV}} \right) \times \phi_0 \times \left[  \int_{2 E_{\tau}^{min}}^{2 E_{\tau}^{max}} d E_\chi \cdot \epsilon_{q}(E_\chi) \cdot \left(\frac{E_\chi}{\text{EeV}}\right)^{- \gamma} + \int_{4 E_{\tau}^{min}}^{4 E_{\tau}^{max}} d E_\chi \cdot \epsilon_{g}(E_\chi) \cdot \left(\frac{E_\chi}{\text{EeV}}\right)^{- \gamma} \right]
\end{equation}
where the limits of integration are determined by the $1\sigma$ range of observed $\tau$ energy. Note that the limits and $\epsilon_{q/g}$ depend on the chord length in consideration. Keeping $\mathcal{N} = 1$, one can obtain the required value of $\phi_0$ for various choices of $\gamma$. The results have been summarised in Table \ref{tab2}. It can be seen that these values are compatible with the upper bounds mentioned in Sec. II. Note that, for $\gamma = 0$, one expects more number of events with shower energies higher than the ones observed by ANITA. Hence, higher values of spectral index is preferred. \\

\begin{table}[h!]
	\centering
		\begin{tabular}{|c|S|S|S|S|S|S|S|S|}
		\hline
		\toprule
		\multirow{2}{*}{} &
		\multicolumn{2}{c|}{$\gamma = 0$} &
		\multicolumn{2}{c|}{$\gamma = 1$} &
		\multicolumn{2}{c|}{$\gamma = 2$}&
		\multicolumn{2}{c|}{$\gamma = 3$} \\
		\cline{2-9}
		& {A} & {B} & {A} & {B} & {A} & {B} & {A} & {B} \\
		\hline
		\midrule
		$\phi_0$&0.19 & 0.41 & 0.31 & 0.71 & 0.37 & 1.04 & 0.33  & 1.30\\
		\bottomrule
		\hline
	\end{tabular}
\caption{\label{tab2} The required value of $\phi_0$ to get $\mathcal{N} = 1$ for various choices of spectral index and chord lengths ($A \equiv 5740$ km and $B \equiv 7210$ km).}
\end{table}

We briefly comment regarding the source of such high energy sterile neutrinos. They can either be produced via the leptoquark interactions, via oscillation of active neutrinos near the source, or via interactions during propagation. If the sterile neutrinos are produced due to oscillation from the active ones, then the flux is proportional to the square of the mixing angle. For large mixing, the cross section will dominated by SM interactions and the sterile neutrino will be significantly attenuated by Earth. For small mixing, albeit the sterile neutrino propagates freely, the incident flux is smaller and constraints from active neutrino flux becomes important. On the other hand, if a flux of active neutrinos encounters large magnetic fields during propagation, it can convert to sterile neutrinos via the transition magnetic dipole moment \cite{Barranco:2012xj}. In this scenario, one anticipates both fluxes to be of the same order of magnitude and offers a lucrative testable explanation. Another possibility is the absorption of active neutrino flux by cosmic sterile neutrino background \cite{Chauhan:2018dkd} or dark matter \cite{Mohanty:2018cmq, Karmakar:2018fno}. In \cite{Heurtier:2019git}, a flux of boosted right handed neutrinos was obtained through decay of dark matter. \\

\section{Conclusion} 
Since the observation of AAEs, many BSM scenarios have been invoked to explain the discrepancy. In this paper we have proposed two models that can significantly enhance the $\tau$ survival probability while simultaneously addressing the flavor anomalies. In the first scenario, we have extended chiral vector leptoquark model which explains $R(D^{(\ast)})$ and $R(K^{(\ast)})$ \cite{Buttazzo:2017ixm} by a sterile neutrino. The cosmogenic UHE neutrinos interact with the matter in Earth and produce a sterile neutrino that propagates freely inside Earth and decays near the surface to a $\tau$. The precise measurement of $Br(B_c \rightarrow \tau \chi)$, which is possible in upcoming B factories, will provide a good test of this model. \\

In the second scenario, a cosmogenic UHE sterile neutrino passes through the Earth almost unattenuated and interacts with the matter in Earth to produce an observable $\tau$. The same interactions and parameters also explain $R(D^{(\ast)})$ anomaly \cite{Ref-DG}. The interesting result is that the distribution of emergent $\tau$ energy peaks in the same regime as observed by ANITA. This model has observable signatures in 300 fb$^{-1}$ LHC searches. \\

In summary, the observation of lepton flavor universality violation and Earth emergent $\tau$ with EeV energy can be explained in a common framework. Moreover, it has testable signatures in upcoming experiments. Future observations by IceCube Gen-II and data from ANITA-IV should be able to shed more light on such BSM hypotheses.


\begin{thebibliography}{99}
	

\bibitem{Gorham:2016zah} 
P.~W.~Gorham {\it et al.} [ANITA Collaboration],
Phys.\ Rev.\ Lett.\  {\bf 117}, no. 7, 071101 (2016)
doi:10.1103/PhysRevLett.117.071101
[arXiv:1603.05218 [astro-ph.HE]].


\bibitem{Gorham:2018ydl} 
P.~W.~Gorham {\it et al.} [ANITA Collaboration],
Phys.\ Rev.\ Lett.\  {\bf 121}, no. 16, 161102 (2018)
doi:10.1103/PhysRevLett.121.161102
[arXiv:1803.05088 [astro-ph.HE]].

\bibitem{Schoorlemmer:2015afa} 
H.~Schoorlemmer {\it et al.},
Astropart.\ Phys.\  {\bf 77}, 32 (2016)
doi:10.1016/j.astropartphys.2016.01.001
[arXiv:1506.05396 [astro-ph.HE]].




\bibitem{Ref-Fox} 
D.~B.~Fox, S.~Sigurdsson, S.~Shandera, P.~Mészáros, K.~Murase, M.~Mostafá and S.~Coutu,
[arXiv:1809.09615 [astro-ph.HE]].

\bibitem{Cherry:2018rxj} 
J.~F.~Cherry and I.~M.~Shoemaker,
arXiv:1802.01611 [hep-ph].


\bibitem{Ref-Huang} 
G.~Y.~Huang,
Phys.\ Rev.\ D {\bf 98}, no. 4, 043019 (2018)
doi:10.1103/PhysRevD.98.043019
[arXiv:1804.05362 [hep-ph]].

\bibitem{Ref-Bhu} 
J.~H.~Collins, P.~S.~Bhupal Dev and Y.~Sui,
arXiv:1810.08479 [hep-ph].




\bibitem{Anchordoqui:2018ssd} 
L.~A.~Anchordoqui and I.~Antoniadis,
arXiv:1812.01520 [hep-ph].

\bibitem{Anchordoqui:2018ucj} 
L.~A.~Anchordoqui, V.~Barger, J.~G.~Learned, D.~Marfatia and T.~J.~Weiler,
LHEP {\bf 1}, no. 1, 13 (2018)
doi:10.31526/LHEP.1.2018.03
[arXiv:1803.11554 [hep-ph]].







\bibitem{Hiller:2003js} 
G.~Hiller and F.~Kruger,
Phys.\ Rev.\ D {\bf 69}, 074020 (2004)
doi:10.1103/PhysRevD.69.074020
[hep-ph/0310219].




\bibitem{Huschle:2015rga} 
M.~Huschle {\it et al.} [Belle Collaboration],
Phys.\ Rev.\ D {\bf 92}, no. 7, 072014 (2015)
doi:10.1103/PhysRevD.92.072014
[arXiv:1507.03233 [hep-ex]].


\bibitem{Hirose:2016wfn} 
S.~Hirose {\it et al.} [Belle Collaboration],
Phys.\ Rev.\ Lett.\  {\bf 118}, no. 21, 211801 (2017)
doi:10.1103/PhysRevLett.118.211801
[arXiv:1612.00529 [hep-ex]].

\bibitem{Aaij:2017deq} 
R.~Aaij {\it et al.} [LHCb Collaboration],
Phys.\ Rev.\ D {\bf 97}, no. 7, 072013 (2018)
doi:10.1103/PhysRevD.97.072013
[arXiv:1711.02505 [hep-ex]].


\bibitem{Aaij:2017vbb} 
R.~Aaij {\it et al.} [LHCb Collaboration],
JHEP {\bf 1708}, 055 (2017)
doi:10.1007/JHEP08(2017)055
[arXiv:1705.05802 [hep-ex]].




\bibitem{Buttazzo:2017ixm} 
D.~Buttazzo, A.~Greljo, G.~Isidori and D.~Marzocca,
JHEP {\bf 1711}, 044 (2017)
doi:10.1007/JHEP11(2017)044
[arXiv:1706.07808 [hep-ph]].

\bibitem{Assad:2017iib} 
N.~Assad, B.~Fornal and B.~Grinstein,
Phys.\ Lett.\ B {\bf 777}, 324 (2018)
[arXiv:1708.06350 [hep-ph]].

\bibitem{Crivellin:2018yvo} 
A.~Crivellin, C.~Greub, D.~Müller and F.~Saturnino,
Phys.\ Rev.\ Lett.\  {\bf 122}, no. 1, 011805 (2019)
doi:10.1103/PhysRevLett.122.011805
[arXiv:1807.02068 [hep-ph]].


\bibitem{Ref-DG} 
A.~Azatov, D.~Barducci, D.~Ghosh, D.~Marzocca and L.~Ubaldi,
JHEP {\bf 1810}, 092 (2018)
doi:10.1007/JHEP10(2018)092
[arXiv:1807.10745 [hep-ph]].

\bibitem{Ref-Bes} 
A.~Angelescu, D.~Bečirević, D.~A.~Faroughy and O.~Sumensari,
JHEP {\bf 1810}, 183 (2018)
[arXiv:1808.08179 [hep-ph]].

\bibitem{Biswas:2018snp} 
A.~Biswas, D.~Kumar Ghosh, N.~Ghosh, A.~Shaw and A.~K.~Swain,
arXiv:1808.04169 [hep-ph].










\bibitem{Bhattacharya:2014wla} 
B.~Bhattacharya, A.~Datta, D.~London and S.~Shivashankara,
Phys.\ Lett.\ B {\bf 742}, 370 (2015)
[arXiv:1412.7164 [hep-ph]].


\bibitem{Bauer:2015knc} 
M.~Bauer and M.~Neubert,
Phys.\ Rev.\ Lett.\  {\bf 116}, no. 14, 141802 (2016)
[arXiv:1511.01900 [hep-ph]].

\bibitem{Fajfer:2015ycq} 
S.~Fajfer and N.~Košnik,
Phys.\ Lett.\ B {\bf 755}, 270 (2016)
[arXiv:1511.06024 [hep-ph]].


\bibitem{Sakaki:2013bfa} 
Y.~Sakaki, M.~Tanaka, A.~Tayduganov and R.~Watanabe,
Phys.\ Rev.\ D {\bf 88}, no. 9, 094012 (2013)
[arXiv:1309.0301 [hep-ph]].

\bibitem{Calibbi:2017qbu} 
L.~Calibbi, A.~Crivellin and T.~Li,
arXiv:1709.00692 [hep-ph].


\bibitem{Crivellin:2017zlb} 
A.~Crivellin, D.~Müller and T.~Ota,
JHEP {\bf 1709}, 040 (2017)
[arXiv:1703.09226 [hep-ph]].


\bibitem{ColuccioLeskow:2016dox} 
E.~Coluccio Leskow, G.~D'Ambrosio, A.~Crivellin and D.~Müller,
Phys.\ Rev.\ D {\bf 95}, no. 5, 055018 (2017)
[arXiv:1612.06858 [hep-ph]].


\bibitem{Blanke:2018sro} 
M.~Blanke and A.~Crivellin,
Phys.\ Rev.\ Lett.\  {\bf 121}, no. 1, 011801 (2018)
[arXiv:1801.07256 [hep-ph]].

\bibitem{Chauhan:2017uil} 
B.~Chauhan and B.~Kindra,
arXiv:1709.09989 [hep-ph].

\bibitem{Dorsner:2016wpm} 
I.~Doršner, S.~Fajfer, A.~Greljo, J.~F.~Kamenik and N.~Košnik,
Phys.\ Rept.\  {\bf 641}, 1 (2016)
[arXiv:1603.04993 [hep-ph]].



\bibitem{Becirevic:2018uab} 
D.~Bečirević, B.~Panes, O.~Sumensari and R.~Zukanovich Funchal,
JHEP {\bf 1806}, 032 (2018)
[arXiv:1803.10112 [hep-ph]].


\bibitem{Clark:2016jgm} 
D.~B.~Clark, E.~Godat and F.~I.~Olness,
Comput.\ Phys.\ Commun.\  {\bf 216}, 126 (2017)
[arXiv:1605.08012 [hep-ph]].


\bibitem{Ball:2017nwa} 
R.~D.~Ball {\it et al.} [NNPDF Collaboration],
Eur.\ Phys.\ J.\ C {\bf 77}, no. 10, 663 (2017)
[arXiv:1706.00428 [hep-ph]].


\bibitem{Ball:2017otu} 
R.~D.~Ball, V.~Bertone, M.~Bonvini, S.~Marzani, J.~Rojo and L.~Rottoli,
Eur.\ Phys.\ J.\ C {\bf 78}, no. 4, 321 (2018)
[arXiv:1710.05935 [hep-ph]].

\bibitem{Tanabashi:2018oca} 
M.~Tanabashi {\it et al.} [Particle Data Group],
Phys.\ Rev.\ D {\bf 98}, no. 3, 030001 (2018).








\bibitem{Alloul:2013bka} 
A.~Alloul, N.~D.~Christensen, C.~Degrande, C.~Duhr and B.~Fuks,
Comput.\ Phys.\ Commun.\  {\bf 185}, 2250 (2014)
[arXiv:1310.1921 [hep-ph]].







\bibitem{Dorsner:2018ynv} 
I.~Doršner and A.~Greljo,
JHEP {\bf 1805}, 126 (2018)
[arXiv:1801.07641 [hep-ph]].

\bibitem{Belyaev:2012qa} 
A.~Belyaev, N.~D.~Christensen and A.~Pukhov,
Comput.\ Phys.\ Commun.\  {\bf 184}, 1729 (2013)
[arXiv:1207.6082 [hep-ph]].

\bibitem{Barranco:2012xj} 
J.~Barranco, O.~G.~Miranda, C.~A.~Moura and A.~Parada,
Phys.\ Lett.\ B {\bf 718}, 26 (2012)
doi:10.1016/j.physletb.2012.10.024
[arXiv:1205.4285 [astro-ph.HE]].

\bibitem{Chauhan:2018dkd} 
B.~Chauhan and S.~Mohanty,
Phys.\ Rev.\ D {\bf 98}, no. 8, 083021 (2018)
doi:10.1103/PhysRevD.98.083021
[arXiv:1808.04774 [hep-ph]].

\bibitem{Mohanty:2018cmq} 
S.~Mohanty, A.~Narang and S.~Sadhukhan,
arXiv:1808.01272 [hep-ph].

\bibitem{Karmakar:2018fno} 
S.~Karmakar, S.~Pandey and S.~Rakshit,
arXiv:1810.04192 [hep-ph].

\bibitem{Heurtier:2019git} 
L.~Heurtier, Y.~Mambrini and M.~Pierre,
arXiv:1902.04584 [hep-ph].

\end{thebibliography}
\end{document}